\documentclass[runningheads]{llncs}
\PassOptionsToPackage{hyphens}{url}
\usepackage{amsmath,amssymb}
\usepackage{hyperref}
\usepackage{cleveref}
\usepackage{graphicx}
\usepackage{xcolor}
\usepackage{booktabs}
\usepackage{subfig}

\title{Collateral Portfolio Optimization in \\ Crypto-Backed Stablecoins}
\titlerunning{Collateral Portfolio Optimization}
\author{Bretislav Hajek$^*$, Dani\"el Reijsbergen$^\dagger$, Anwitaman Datta$^\dagger$, and Jussi Keppo$^*$}
\institute{$^*$National University of Singapore, Singapore \\ $^\dagger$Nanyang Technological University, Singapore}
\authorrunning{Bretislav Hajek, Dani\"el Reijsbergen, Anwitaman Datta, and Jussi Keppo}

\begin{document}

\newcommand{\name}{\textsc{CroCoDai}}
\newcommand{\rda}[1]{\textcolor{black}{#1}}

\maketitle

\begin{abstract}
Stablecoins -- crypto tokens whose value is pegged to a real-world asset such as the US Dollar -- are an important component of the DeFi ecosystem as they mitigate the impact of token price volatility. In crypto-backed stablecoins, the peg is founded on the guarantee that in case of system shutdown, each stablecoin can be exchanged for a basket of other crypto tokens worth approximately its nominal value. However, price fluctuations that affect the collateral tokens may cause this guarantee to be invalidated. In this work, we investigate the impact of the collateral portfolio's composition on the resilience to this type of catastrophic event. \rda{For stablecoins whose developers maintain a significant portion of the collateral (e.g., MakerDAO's Dai), we propose two portfolio optimization} methods, based on convex optimization and (semi)variance minimization, that account for the correlation between the various token prices. We compare the optimal portfolios to the historical evolution of Dai's collateral portfolio, and to aid reproducibility, we have made our data and code publicly available.

\end{abstract}

\section{Introduction}
\label{sec:introduction}
Decentralized Finance (DeFi) promises to improve on traditional finance (TradFi) in areas such as transparency, efficiency, and censorship resistance \cite{werner2022sok}. However, the early days of DeFi have been characterized by intense price volatility -- e.g., after an initial boom in 2021, 
 the prices of key DeFi tokens such as Ether (ETH)  and \rda{Wrapped} Bitcoin (\rda{WBTC}) fell more than 60\% in a period of two months after April 2022.\footnote{\url{https://coinmarketcap.com/}} 
High price volatility limits the suitability of crypto tokens as a medium of exchange -- e.g., because token prices change considerably while a trade or auction is ongoing -- and (short-term) store of value, which may hamper the growth of DeFi. \textit{Stablecoins}, whose value is pegged to a real-world asset, promise to mitigate this volatility. However, stablecoins themselves are not invulnerable to losing their peg, as witnessed by the sudden collapse of the NuBits \cite{moin2020sok} and Terra/Luna \cite{briola2023anatomy} stablecoins. As such, it is vital that stablecoin risks are understood and that the insights are leveraged to mitigate volatility risks.

A variety of different stablecoin designs have emerged, \rda{which can be grouped into three broad categories \cite{klages2020stablecoins,moin2020sok}.} The first category is that of \textit{fiat-backed stablecoins}, which are backed by an equivalently valued amount of fiat collateral. The second category is that of crypto-collateral-backed (\textit{crypto-backed} for brevity) stablecoins, which are backed by regular -- i.e., volatile -- crypto tokens stored in smart contract \textit{vaults} as collateral. The central guarantee in crypto-backed stablecoins is that in case of system shutdown, each stablecoin can be redeemed for a basket of crypto tokens with equivalent value. \rda{The third category is that of \textit{algorithmic stablecoins}, which do not have collateral and maintain their peg by adjusting the supply of tokens based on fluctuating exchange rates.} Each category has its drawbacks: fiat-backed stablecoins rely on the same trusted intermediaries that enable TradFi, and therefore share some of TradFi’s drawbacks, e.g., reliance on trusted parties, potential of censorship, and inefficient creation of new tokens. \rda{Meanwhile, algorithmic stablecoins have nothing to support them in case of a death spiral \cite{briola2023anatomy,moin2020sok}, and the examples of NuBits \cite{moin2020sok} and Terra/Luna \cite{briola2023anatomy} demonstrate that this design is vulnerable to collapse. For crypto-backed stablecoins, the main risk is that price fluctuations cause the value of the collateral to drop below the number of stablecoins in circulation, thus violating the stablecoin's central guarantee.\footnote{\rda{Although we are unaware of examples of crypto-backed stablecoins that have collapsed entirely due to price fluctuations, there have been instances in which they may have come close (e.g., Dai’s `Black Thursday' event in March 2020 \cite{kjaer2021empirical}).}}
To mitigate this risk,} this design requires \textit{overcollateralization}, i.e., that the ratio of the value of the collateral to the number of issued stablecoins in each vault always exceeds some threshold $\theta > 1$. The notoriously high price volatility of crypto tokens necessitates high thresholds in practice, which can be reduced by 1)\ adding fiat-backed stablecoins and real-world assets, leading to \textit{hybrid} fiat/crypto-backed stablecoins, and 2)\ minimizing the crypto portion's risk.

In this work, we investigate MakerDAO's Dai (DAI) stablecoin \cite{daiwhitepaper}, which is underpinned by the Dai Stablecoin System (DSS). While originally a `pure' crypto-backed stablecoin, the general bust in the DeFi market in 2022 has led MakerDAO to explore alternative means of collateral, including native support for certain fiat-backed stablecoins (\textit{peg stability modules}) and real-world assets maintained on behalf of the Maker Foundation \cite{yahoo2023makerdao,dlnews2023makerdao}. In particular, regular crypto tokens have accounted for less than 40\% of Dai's total collateral since late 2023, \rda{while nearly 60\% being held in the form of MakerDAO's real-world assets. As such, Dai can be seen as a prominent example of a hybrid fiat/crypto-backed stablecoin}. However, the decision to incorporate real-world assets has not been entirely uncontroversial, with several members of MakerDAO's core engineering team resigning in protest \cite{dlnews2023endgame}. MakerDAO has floated a long-term plan of a maximum contribution of real-world assets to Dai's collateral of 25\% \cite{yahoo2023makerdao}. As a consequence, the crypto portion is scheduled to increase, which makes choosing its composition a pertinent research question \rda{because MakerDAO has the ability to transfer its sizeable portion of real-world assets into underrepresented tokens}. The risks associated with the crypto portion  can be estimated through simulation, which has previously been applied to investigate the risks of individual Dai vaults -- i.e., \textit{liquidation} if the overcollateralization drops below $\theta$ -- by academic researchers \cite{bhat2021simulating,kirillov2022stablesims,chaleenutthawut2024loan} and by crypto research firms \cite{bluhm2024real,osolnik2022maker}. In this work, we take a broader view and focus on the collateralization of the stablecoin as a \textit{whole} instead of individual vaults. We use convex optimization to find portfolios with minimal combined variance, taking into account the correlation between token prices. We also present a procedure that uses \textit{semi}variance minimization, which exploits the knowledge that upward volatility (i.e., prices increasing) is not as harmful to Dai's guarantee as downward volatility.  Our method is \textit{dynamic} in the sense that it is based on recent price data, and that it prescribes how to modify a collateral portfolio based on evolving market conditions. We compare the optimal portfolios to the historical evolution and find that Dai's resilience would be greater if the contribution of Ether were reduced in favor of (wrapped) Bitcoin. Finally, we have made all our data and code public via \url{https://github.com/ntublockchain/dai-collateral-data}.

\subsubsection*{Contributions.}

In summary, our main contributions are as follows.

\begin{itemize}
    \item We use variance and semivariance minimization to determine optimal crypto collateral portfolios (\Cref{sec:portfolio_optimization}). Our method is dynamic, i.e., based on evolving price data and it can be used for continuous decision support.
    \item We compare the optimized portfolios to the historical evolution of Dai collateral and find that its resilience could be augmented through the inclusion of more WBTC tokens (\Cref{sec:results}).
    \item Our code and dataset have been made publicly accessible.
\end{itemize}

\subsubsection*{Outline.} This work is structured as follows. We discuss the preliminaries and related work in \Cref{sec:background}. We discuss our portfolio optimization method in \Cref{sec:portfolio_optimization}, and the details of the empirical data collection for the Dai stablecoin in \Cref{sec:empirical_data}. We compare the optimal portfolios to the historical evolution of Dai collateral in \Cref{sec:results}, and \Cref{sec:conclusions} concludes the paper.

\section{Preliminaries \& Related Work}
\label{sec:background}

\subsection{DeFi Concepts}
\label{sec:defi_concepts}

We assume that the reader is familiar with basic DeFi concepts such a blockchains, transactions, smart contracts, and Ethereum (see, e.g., \cite{buterin2014next,narayanan2016bitcoin,werner2022sok} for more details). However, we provide a quick summary of other DeFi concepts that appear later in this work to clarify terminology.

\textbf{ERC-20 Standard. } Smart contracts can be used to define local token types beyond a blockchain's native token. In Ethereum, this process is facilitated by the ERC-20 token standard.
ERC-20-compliant smart contracts maintain a common set of functions for tokens, e.g., transfers and balance inquiries.

\textbf{Decentralized Exchanges (DEXs). } Tokens that are local to the same blockchain can be exchanged without the need for an intermediary by using a DEX that operates an 
\textit{Automated Market Maker} (AMM) protocol \cite{xu2023sok}. The biggest example of a DEX is Uniswap.\footnote{\url{https://uniswap.org/}} As of V2, a Uniswap DEX on the Ethereum blockchain consists of a separate smart contract for any pair of ERC-20 tokens. Users can withdraw tokens of one type from the pool in return for a number of tokens of the other type as described by the AMM. To provide liquidity, users can deposit tokens of both types to earn \textit{Liquidity Pool} (LP) tokens, which entitle holders to a portion of the fees collected by the DEX.

\textbf{Wrapped Tokens. } A token from one blockchain can exist on another blockchain in the form of a \textit{wrapped} token -- e.g., Wrapped Bitcoin (WBTC), which is an ERC-20 token on Ethereum. Wrapped tokens are typically backed by matching collateral on the original blockchain -- this collateral can be controlled by trusted custodians (e.g., WBTC) or through a protocol with overcollateralized  vaults (e.g., \textsc{Xclaim} \cite{zamyatin2019xclaim}).

\textbf{Event Logs. } Smart contracts in Ethereum can emit events defined by the smart contract code. Events consist of up to 4 32-byte `words' and are stored separately in the block (i.e., blocks have a separate Patricia-Merkle tree for logs). Logs allow clients to efficiently query the state of contracts after a transaction.

\subsection{Related Work}

\textbf{Stablecoin Models \& Surveys. } High-level descriptions of the principles of crypto-backed stablecoins can be found in various survey papers \cite{klages2020stablecoins,werner2022sok,ante2023systematic}. Werner et al.\ \cite{werner2022sok} provide a general overview of DeFi that also includes a brief description of stablecoins. Klages-Mundt et al.\ \cite{klages2020stablecoins} provide a high-level overview of research challenges in stablecoins, including a classification of stablecoin designs (custodial vs.\ non-custodial), a discussion of price and capital structure models from the scientific literature, and agent incentives. Ante et al.\ \cite{ante2023systematic} provide a systematic literature review of 22 papers that perform empirical evaluations of stablecoins, of which the majority focus on fiat-backed stablecoins. 

Salehi et al.\ propose the formalism of red-black coins \cite{salehi2021red}, which combine stablecoins with another token type that incurs added volatility for a fee -- the DSS can be seen as an extension of this concept. 
They also investigate the removal of liquidations from the DSS. 
Cao et al.\ \cite{cao2022designing} propose two methods that derive stablecoins and leveraged investment instruments from regular tokens. Both works illustrate their methods using simulations of geometric Brownian motion (the latter via a jump-diffusion model).

\textbf{Stability of Dai Vaults. } The stability of Dai vaults is an active area of research both among crypto firms \cite{bluhm2024real,osolnik2022maker} and in academia \cite{bhat2021simulating,kirillov2022stablesims,kjaer2021empirical,chaleenutthawut2024loan}. Bluhm et al.\ propose the CALM framework \cite{bluhm2024real} to evaluate the risks of crypto collateral -- they consider duration risk, credit risk, market risk, and operational risk. The market risk in CALM is equivalent to the risk studied in this work, i.e., a loss of value of the crypto collateral due to price fluctuations. To evaluate market risk, the authors cite work by Block Analitica \cite{osolnik2022maker} on the Maker Risk Dashboard.\footnote{\url{https://maker.blockanalitica.com/simulations/risk-model/}} The latter uses simulations to recommend system-level parameters for DSS vault types, e.g., the savings rate and debt ceiling. While similar in aim to our work, the simulations have fixed parameters (e.g., 50\% price drops for ETH) that do not depend on recent historical data and which do not take the correlation between token price changes into account.

Simulation-based estimation of the credit risk in individual vaults also appears in \cite{bhat2021simulating} and \cite{kirillov2022stablesims}. The former work focuses on modeling the price of Dai given a population of investors with varying risk tolerance profiles, whereas the latter focuses on simulating the liquidation auction mechanism. Kj\"aer et al.\ study Dai's liquidation auction mechanism during the Black Thursday event of March 2020 \cite{kjaer2021empirical}, during which a large number of vaults were liquidated. Chaleenutthawut et al.\ provide a dataset of interactions with the ETH-A vault type \cite{chaleenutthawut2024loan}. This can be used to study the evolution of the collateral in individual vaults over time. Liquidation risks for individual vaults are also computed using Brownian motion.

\textbf{Others. } Like our work, Balance \cite{harz2019balance} aims to reduce the overcollateralization rates in crypto deposits (Dai's collateral is included as an example) through portfolio optimization, however the focus is on attacks rather than price fluctuations. CroCoDai \cite{reijsbergen2023crocodai} aims to improve the resilience of crypto-backed stablecoins by allowing native support of collateral on multiple chains without the need for (expensive) cross-chain transfers of wrapped tokens.

\subsection{\rda{Existing Stablecoins}}

\rda{A multitude of stablecoin designs have emerged in practice, and we use this section to highlight some of the most prominent designs. We focus on crypto-backed stablecoins as they are the main focus of this paper.}

\rda{\textbf{Crypto-backed stablecoins}: Dai is the most prominent (by market value) stablecoin with a crypto-backed component, although it is more accurately referred to as a hybrid fiat/crypto-backed stablecoin because around 60\% of its collateral consists of real-world assets held by the Maker Foundation. Examples of \textit{pure} crypto-backed stablecoins include Liquity USD (LUSD) and Reflexer's RAI, which are both backed entirely by ETH vaults.}

\rda{The Frax \cite{kazemian2022frax} platform supports a hybrid crypto-backed/algorithmic stablecoin (FRAX). FRAX, or any Frax-like coin, is a fully crypto-backed stablecoin upon its inception -- users can only mint new coins by depositing other ERC20 tokens such as USDT (Tether), BUSD, sUSD, USDC, and Dai. However, once FRAX has achieved significant usage, its algorithmic component -- in which tokens are minted or burned when the currency loses its peg -- becomes more prominent. Celo \cite{clabs2019celo} is a layer-2 Ethereum platform which supports its own stablecoin (CUSD), which is backed by a combination of fiat and crypto assets including Celo's ERC-20 token (CELO) and other cryptocurrencies such as Bitcoin and Ether.}

\rda{\textbf{Other stablecoins}: The most prominent examples of fiat-backed stablecoins include Tether (USDT) and Circle's USD Coin (USDC). As mentioned in \cite{jiang2023decentralized}, prominent examples of pure (non-collateralized) algorithmic stablecoins include Ampleforth (AMPL), and the now-defunct Terra (UST), NuBits (USNBT), and BaseCoin (BAB). The large number of collapsed algorithmic stablecoins has called the viability of this approach in doubt \cite{moin2020sok}.}

\section{Portfolio Optimization}
\label{sec:portfolio_optimization}
In this section, we discuss portfolio optimization in crypto-backed stablecoins using (semi)variance minimization. As our approach is not specific to the Dai stablecoin, we will keep our discussion high-level and illustrate our approach using tokens beyond the Ethereum ecosystem.
In the following, we first focus on obtaining the optimal portfolio in an idealized setting in which protocol engineers have perfect control of its composition in \Cref{ec:finding_portfolio}, and then discuss in \Cref{sec:enforcing_portfolio} what options protocol engineers have in practice to enforce it.

\subsection{Obtaining Optimal Portfolios}
\label{ec:finding_portfolio}

The primary way to increase a crypto-backed stablecoin's resilience to price shocks is to obtain a portfolio that minimizes volatility, especially downward movements. Large, sudden price drops can result in a liquidation of users' deposits and potentially destabilize the entire protocol.
Even though positive returns are beneficial for a portfolio's stability, the mean returns vary significantly over the observed years and are often close to zero in the medium-term. By contrast, we have observed from our data that variance values are more persistent over time. Therefore, the mean returns are not our primary optimization criterion, and we focus first on minimizing the volatility while considering the (possibly high) correlation between token prices. 

Using principles from mean-variance analysis, the optimal portfolio composition is found by solving the convex optimization problem defined as
\begin{equation}
\begin{array}{rcl}
\displaystyle \min_{\boldsymbol{a}} & \boldsymbol{a}^{T} C \boldsymbol{a}, & \\[0.2cm]
\text{s.t.} & \displaystyle  \sum_{i=1}^M a_i = 1, & \\[0.2cm] 
& \displaystyle  0 \leq a_i \leq \lambda_i \;&\; i = 1,\ldots,M,
\end{array}
\label{eq:convex_optimization}
\end{equation}
where $M$ denotes the number of included tokens, $a_i$ the fraction of the portfolio that is invested in token $i \in \{1,\ldots,M\}$, $\boldsymbol{a}$ the vector $(a_1,\ldots,a_M)$, and $C$ the $M \times M$ covariance matrix of the prices of the included cryptocurrencies. 
Here, the prices are assumed to be random variables whose probability distribution has the same (finite) mean and variance over short time windows. 
The bounds \mbox{$\lambda_i \in (0,1)$} prevent portfolios that are dominated by a single token and therefore more prone to black-swan events (e.g., the collapse of the FTX token \cite{jalan2023systemic}). 

\begin{figure}[t]
    \centering
    \includegraphics[width=0.9\linewidth]{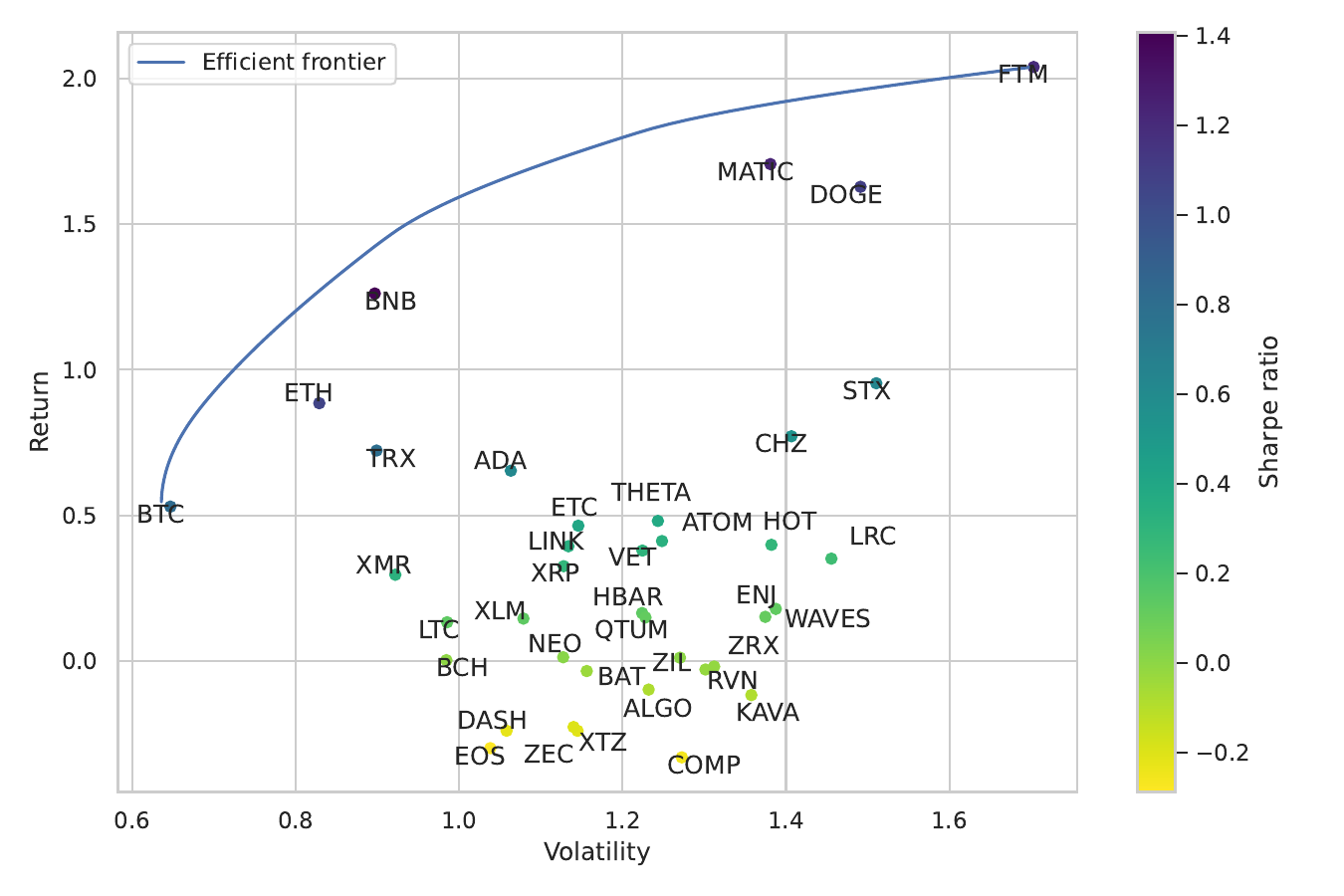}
    \caption{Efficient frontier representing optimal portfolios consisting of selected cryptocurrencies for different values of expected return.}
    \label{fig:ef_frontier}
\end{figure}

Regarding the mean returns, \Cref{fig:ef_frontier} visualizes the trade-off between price (normalized) volatility and expected returns for different token types, as indicated by their symbol (e.g., BTC for Bitcoin). We observe an efficient frontier of optimal portfolios, i.e., the portfolios with the highest return for a given level of risk. 
In \Cref{fig:ef_frontier}, we have also displayed the tokens' Sharpe ratios \cite{sharpe1994sharpe} -- higher values indicate higher rewards for the same volatility. We observe that BTC has the lowest volatility in our dataset, but not the highest returns. By contrast, the Fantom token (FTM) has the highest returns, but at the cost of high volatility.

\textbf{Minimum Semivariance.}
Volatility minimization as discussed above treats downward and upward price movements similarly. However, the risk of a stablecoin crash comes only from
downward movement.
This observation can be captured through the notion of \textit{semivariance}.\footnote{The semivariance of a random variable $X$ is defined as $\mathbb{E}[(X - \mathbb{E}[X])^2 \textbf{1}(X \leq \mathbb{E}[X])]$ where $\textbf{1}$ is the standard indicator function.} 
Computation methods exist for finding optimal portfolios that minimize this risk (called mean-semivariance optimization). The main downside of this method is higher computational complexity, which can become an obstacle when considering a large number of assets over long periods \cite{semivariance}.
For the evaluation in \Cref{sec:results}, we will consider optimal portfolios under both minimum variance and minimum semivariance conditions. 

\textbf{Token Selection.}
Before running the optimization itself, the set of tokens used for optimization has to be selected. The volatility of returns is not the only factor which needs to be considered during this selection. Tokens included in the portfolio must be trusted by users and provide enough liquidity to estimate their price accurately. These criteria imply that we must limit the set of optimized tokens to the top 100 most popular cryptocurrencies. Further, we reduce this list by removing tokens released within the last 3 years because tokens that have weathered the crypto market bust of 2022 are less likely to experience a sudden collapse due to poor foundations (e.g., bugs or serious protocol weaknesses).

If we use convex optimization across this set, we obtain a portfolio entirely consisting of other stablecoins. Although these fiat-backed stablecoins have naturally low price volatility, they come with their own drawbacks as mentioned in \Cref{sec:introduction}. For that reason, we consider only cryptocurrencies which are not backed by real-world assets. This leads to the most challenging portfolio setting, as these cryptocurrencies are volatile and often highly correlated.

These steps result in a final set of 38 tokens, which are used for the optimization. To ensure greater diversification of the portfolio, we limit the maximum contribution of a single asset by $\lambda_i = 0.2$ for all $i$. Minimizing volatility and semivariance results in portfolios denoted by A-Vol and A-Sem, respectively. We will compare these portfolios in \Cref{sec:results} to portfolios that consist only of ERC-20 tokens, to gain insight into the magnitude of the improvements over the historical portfolios, and to explore the potential benefits (in terms of collateral resilience) of including a wider range of wrapped tokens on Ethereum.

\textbf{Portfolio Evaluation.}
Once the different portfolios have been determined (i.e., Ethereum-only or not), we can compare them in terms of the main metrics that the portfolios are minimized over, i.e., the annual volatility and annual \textit{semideviation} -- i.e., the square root of the semivariance. In addition, we use stochastic simulations to gain a further understanding of the risk involved.
As mentioned before, the main risk comes from a sudden price drop, which may result in the liquidation of the user's deposit. If the price drop is very big, many user vaults can go into liquidation, and the stablecoin may become unpegged. 
Therefore, we evaluate portfolios in terms of the risk of the total overcollateralization ratio dropping below the expected ratio.

The main simulation parameters are the initial overcollateralization ratio, $\gamma$, and the required overcollateralization ratio $\theta$ that has to be maintained at all times. We denote the portfolio's value at time $t$ by $v(t) = \sum_{i=1}^M n_i p_i(t)$, where $n_i$, $i = 1,\ldots,M$, is the number of collateral tokens of type $i$ and $p_i(t)$ their price at time $t$. We assume that $n_i$ remains constant for the duration of the simulation. We note that $n_i = a_i \sum_j n_j$, where $a_i$ is the fraction defined previously, and that the value of the issued stablecoins is given by $v(0)/\gamma$. The stablecoin's portfolio fails if its value drops below the required threshold, i.e., (after some rearranging)
\begin{align}
    \frac{v(t)}{v(0)} \geq \frac{\theta}{\gamma}.
\label{eq:failure_condition}
\end{align}
Portfolios are evaluated using two main strategies: simulations using historical data and using geometric Brownian motion \cite{salehi2021red,cao2022designing}. During historical simulations, periods of historical price data are sampled at random and used for the evaluation of portfolios. Similarly, for geometric Brownian motion, we randomly sample time periods to estimate the mean and variance of the returns and the covariance between individual tokens. These parameters are then used to simulate the prices and evaluate the portfolio. Due to space limitations, we assume an initial overcollateralization $\gamma=2$ and requirement $\theta=1.5$ throughout all simulations. For each simulation result (see \Cref{tab:portfolio_comparison} in \Cref{sec:results}), we use 10\,000 runs to estimate the probability that a portfolio fails over a 365-day period.

There are other mechanisms involved in a crypto-backed stablecoin that we do not consider in the simulations. The stability fee, which is a form of interest paid by tokens holders, plays an important role as it incentivizes users to lock or withdraw their assets. However, we have found that the stability fees are typically too low (i.e., $5$--$10\%$ annually) to have a major influence on our simulations during short time periods, so we not consider them further. \rda{In particular, the main effect that we witnessed in our simulations is that stability fees effectively increase the overcollateralization level over time, as vault owners are required to gradually provide added collateral. However, this effect is only noticeable over longer time periods for realistic fee levels. Hence, the protocol engineers' main goal is still to choose a sufficiently high $\theta$ for the stablecoin to survive short-term price crashes, even if stability fees may boost long-term overcollateralization.}

Another factor that we do not consider for the simulations is the cost associated with transaction fees. The transaction fees are paid by users and do not affect the portfolio value directly. However, large transactions fees may discourage users from using the protocol, hence reducing liquidity. This further justifies the focus on evaluating the stability of the portfolio itself: lower risk of failure, as described by \eqref{eq:failure_condition}, potentially decreases the number of withdrawals and liquidation actions, resulting in lower running costs incurred by users.

\subsection{Enforcing Optimal Portfolios}
\label{sec:enforcing_portfolio}

Given the knowledge of an optimized portfolio, the next question is how protocol engineers can enforce it in practice. \rda{In \textit{hybrid} crypto-backed stablecoin systems,} the protocol developers can use their own funds to purchase underrepresented tokens and deposit it as collateral. For example, as we will discuss in more detail in \Cref{sec:results}, MakerDAO already controls nearly 60\% of all of Dai's collateral through real-world assets, in addition to an unknown number of crypto vaults. 

In a \rda{\textit{pure}} crypto-backed stablecoin system, collateral is, in principle, deposited primarily by a large multitude of users who have their own incentives -- for example, as of Feb.\ 2024, the ETH-A vault type in the DSS had 1011 individual vaults,\footnote{\url{https://maker.blockanalitica.com/vault-types/}} even if some of these are potentially controlled by the same entity. \rda{In this case, protocol engineers have only limited options to enforce portfolios.}
One way is for the protocol engineers to set stringent debt ceilings (similar to the $\lambda_i$ parameters of \eqref{eq:convex_optimization}) for overrepresented tokens. However, a downside is that this may discourage token holders from depositing, thus reducing the total value of the collateral and therefore the maximum number of stablecoin tokens. Another method could be for the protocol engineers to subsidize the conversion of overrepresented into underrepresented currencies by paying for the transaction and DEX fees through a smart contract. 
\rda{The difficulty of enforcing optimal portfolios in pure crypto-backed stablecoins is a limitation of the present work, and research into alternative methods of enforcing the portfolio would be an interesting direction for future work. We do emphasize that variance and semivariance optimization both involve minimizing a convex objective function, so even if an optimal portfolio cannot be enforced precisely, it is, in principle, always advantageous to be as `close' to it as possible. }

\section{Empirical DSS Data}
\label{sec:empirical_data}
In this section, we discuss the collection of empirical data from the DSS, the most prominent crypto-backed stablecoin system -- this data allows us to compare the optimal portfolios to historical portfolios in \Cref{sec:results}. We discuss the 4 types of DSS collateral in \Cref{sec:dss_collateral}, the vault structure in \Cref{sec:vault_structure}, and the data collection methods in \Cref{sec:data_collection}.

\subsection{Types of DSS Collateral}
\label{sec:dss_collateral}

As of early 2024, the DSS has four main types of collateral: 1. ERC-20 tokens, 2. ERC-20 liquidity pool tokens, 3. peg stability modules, and 4. real-world assets.

\textbf{1.\ ERC-20 Tokens.}
Depositing Ethereum-supported tokens as collateral is the oldest supported method of creating Dai tokens, going back to the Dai white paper \cite{daiwhitepaper}. 
The exact procedure to create Dai tokens is as follows. First, the user creates a Collateralized Debt Position (CDP) through a call to a vault contract. Next, the user sends the tokens to the contract to fund the CDP. This allows the user to create and withdraw Dai tokens, under the condition that, at all times, the ratio of the value of the collateral to the number of created Dai tokens exceeds the overcollateralization ratio $\theta$. Users who have created Dai gradually pay interest over their position -- this is called the \textit{stability fee}. Token prices are periodically updated through \textit{price oracles}. In the following, we will refer to an individual CDP as a \textit{vault}.\footnote{This is consistent, with, e.g., the Maker Risk Dashboard and \cite{chaleenutthawut2024loan}.} Vaults are grouped by their \textit{vault type}, with different parameter choices per type, including 1.\ the ERC-20 token type, 2.\ the stability fee, 3.\ the overcollateralization ratio, and 4.\ the \textit{debt ceiling}, which is the maximum Dai value that can be created using the vault type's collateral. 

If, due to the stability fee or a price decrease of the collateral tokens, the CDP is not sufficiently overcollateralized, then it can be \textit{liquidated} -- an auction is started, the remaining collateral is sold to to highest bidder(s), and the remaining collateral (if any) is returned to the user. 

\textbf{2.\ ERC-20 Liquidity Pool (LP) Tokens. } As mentioned in \Cref{sec:defi_concepts}, Uniswap DEXs may issue LP tokens to liquidity providers -- in Uniswap V2, such tokens are themselves ERC-20-compliant although they are typically not listed on major exchanges. 
LP vaults differ from regular ERC-20 vaults in the sense that price information about the tokens in not provided by oracles, but inferred from the pool itself using a formula that is beyond the scope of this paper (see, e.g, the  contract code for the DAI/ETH Uniswap LP token oracle \cite{lptokenoracle}).

\textbf{3.\ Peg Stability Modules (PSMs).}
PSMs are a special type of vault that allow for the instant creation of Dai stablecoin by depositing other stablecoins. PSMs do not maintain CDPs, and the created Dai are not subject to stability fees. Any user can deposit Dai in the PSM to withdraw deposited stablecoins if available. PSMs help the DSS maintain its USD peg by providing an alternative route for token holders to buy or sell DAI tokens for \$1 even if counterparties cannot be found on traditional exchanges.

\textbf{4.\ Real World Assets (RWAs). }
RWAs are portfolios held on behalf of MakerDAO by trusted third parties. As of early 2024, the largest RWAs are held by Monetalis, Coinbase, and BlockTower, among 15 RWA vault types in total. 

\subsection{Vault Structure}
\label{sec:vault_structure}

\begin{figure}[t]
    \centering
    \includegraphics[width=0.95\linewidth]{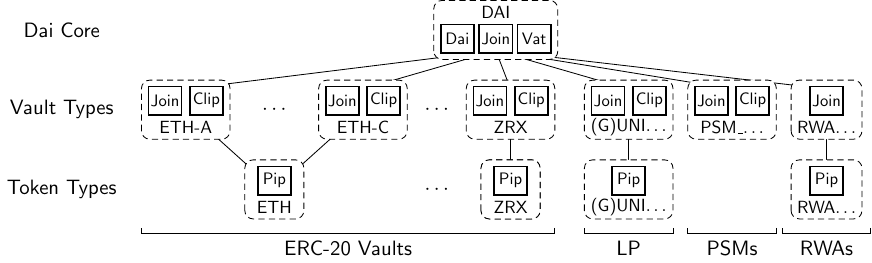}
    \caption{Architecture of collateral-related smart contracts in the DSS. Squares indicate contracts and lines indicate relationships between contracts.}
    \label{dai_contract_structure}
\end{figure}

Each type of vault is implemented in the DSS through multiple smart contracts. The contract types that are the most relevant to this work are the \texttt{Join} contracts, which allow for the submission and withdrawal of collateral, and \texttt{Pip} contracts, which process price information. Tangential to our work are \texttt{Clip} contracts, which handle liquidations, and the \texttt{Vat} contract which contains the code for CDP manipulation (i.e., creating Dai from deposited collateral). Finally, the \texttt{Dai} contract implements the Dai stablecoin as an ERC-20 token. The \texttt{Join} and \texttt{Pip} contracts are particularly relevant to our work because they emit events that represent collateral changes and price updates, thus allowing us to reconstruct the historical evolution of DSS collateral. \rda{A single token type may have multiple vault types -- e.g., ETH has three vault types (ETH-A, ETH-B, and ETH-C) which differ in terms of stability fees and required overcollateralization.}

A visual representation of the relationship between these contract types can be found in \Cref{dai_contract_structure}. At the top is the \texttt{Dai} contract, and each vault type has its own unique \texttt{Join} contract (\texttt{GemJoin}) to provide an interface to \texttt{Dai}. The Dai contract also has its own \texttt{Join} (\texttt{DaiJoin}) contract to enable the withdrawal of Dai stablecoins from a CDP. Each vault type also has its own \texttt{Clip} contract to handle liquidation auctions (except RWAs, which cannot be liquidated), and interacts with a \texttt{Pip} contract that provides price information for its token type -- vault types that share a token type also share the same \texttt{Pip} contract.
For ERC20 and LP tokens, \texttt{Pip} contracts return the median of multiple price feeds. PSMs have no \texttt{Pip} contracts as they are assumed to be 1-for-1 exchangeable with Dai. Finally, RWAs have \texttt{Join} and \texttt{Pip} contracts, but no \texttt{Clip} contract for collateral auctions. RWAs typically feature a single creation of a single batch of $10^{18}$ RWA ``tokens'', and the size of the deposit is updated through the \texttt{Pip} contract.\footnote{For example, the 6s Capital asset (RWA001) was increased from an initial $\$$1,060 on 9 March 2021 to nearly $\$$1.6 million on 27 August 2021 through a ``price'' update.}

\subsection{Data Collection}
\label{sec:data_collection}

To enable our historical comparison, we have collected both vault data (how many tokens were held for each vault type over time) and price data (how much each token type was worth in USD) for the period spanning from the first submission of collateral  (Nov.\ 2019) to the time of writing (Feb.\ 2024).

\textbf{Vault Data. }
After each successful addition or withdrawal of collateral to a \texttt{Join} contract, an event that encapsulates the change is emitted as a 32-byte integer. The full log of events in Ethereum is publicly available a dataset on Google BigQuery \cite{bigquery}.
We obtained the Ethereum addresses of all relevant contracts using the list of contract addresses associated with the Dai stablecoin that can be found on MakerDAO's change log \cite{makeraddresses}.
The resulting collateral compositions for each vault type were validated using Dai Stats \cite{daistats}, which provides snapshots of the current state of  collateral in the DSS.
We obtained vault data for all 59 vault types used for the DSS (by contrast, \cite{chaleenutthawut2024loan} only focus on a single type, ETH-A). 

\textbf{Price Data. }
Price data for ERC-20 tokens can be obtained from Binance via its API \cite{binanceapi}. For LP tokens and RWAs, price information can be obtained from events emitted  by the \texttt{Pip} contracts. We note that the \texttt{Pip} contracts for the other vault types can also be used to obtain price data, although we have not found major discrepancies when comparing the two.

\begin{figure}[t]
    \centering
    \subfloat[][Absolute Contribution]{ \includegraphics[width=\linewidth]{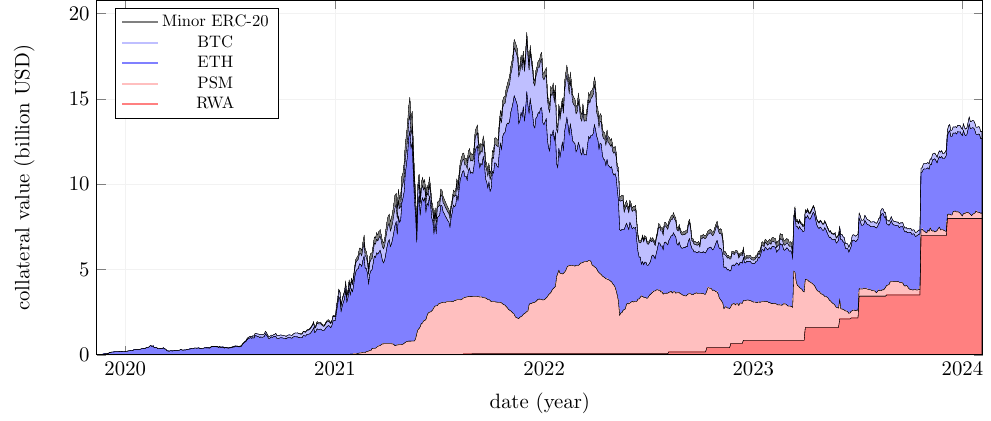}\label{fig:area_plot_hist_abs}} \\
   \subfloat[][Relative Contribution]{ \includegraphics[width=\linewidth]{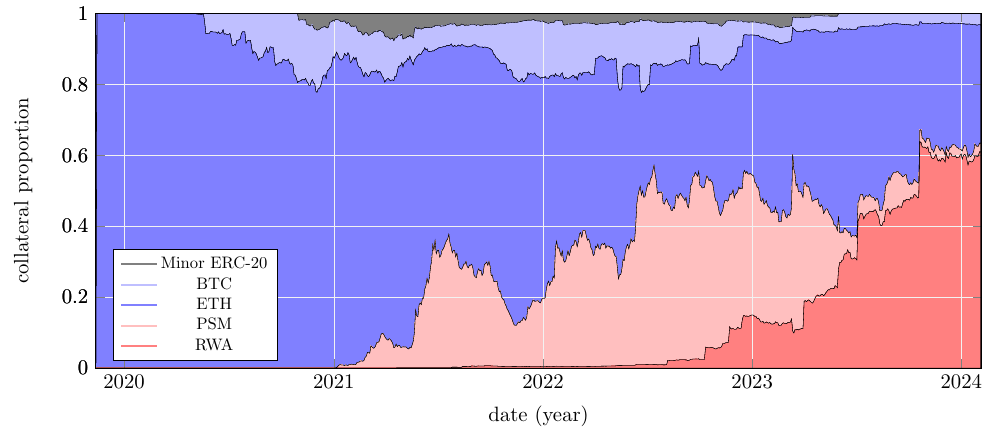}\label{fig:area_plot_hist_rel}}
    \caption{Evolution of the relative (top) and absolute (bottom) contributions of the various collateral types among the total DSS collateral (measured in USD).}
    \label{fig:area_plot_hist}
\end{figure}

\section{Analysis}
\label{sec:results}

\textbf{Historical Evolution. }
\Cref{fig:area_plot_hist} depicts the evolution of the absolute and relative contribution of the various vault type categories. In these figures, the ERC-20 vault category has been subdivided into Ether (ETH), (wrapped) Bitcoin (BTC), and minor tokens. The contribution of  LP tokens was not large enough to be visible. PSMs and RWAs remain as separate categories.

From \Cref{fig:area_plot_hist_abs}, we observe that the total value of the collateral is close to the heights achieved during the 2021-2022 DeFi boom, although this is predominantly due to the increase in PSM and RWA collateral. A sharp rise in the value of the RWA assets is visible in late 2023 -- this corresponds to the addition of $\$$3.6 billion of RWA007 (Monetalis) and RWA015 (BlockTower) collateral in the same Ethereum block (18385588) on 19 October 2023.
We also observe that before 2021, DSS collateral entirely almost consisted of ERC-20 tokens. PSMs (particularly USDC) had a significant contribution until early 2023, after which their share declined. It is unclear why this decline occurred -- one possibility is that a large fraction of the deposited stablecoins in the PSMs were controlled by MakerDAO, and that they gradually converted them into RWAs. As of early 2024, RWAs are the largest contributor to Dai's collateral portfolio with a combined share of approximately $60\%$ (see \Cref{fig:area_plot_hist_rel}).

\begin{figure}[p]
    \centering
   \subfloat[][Optimal Portfolio: 30-Day Window]{ \includegraphics[width=1.0\linewidth]{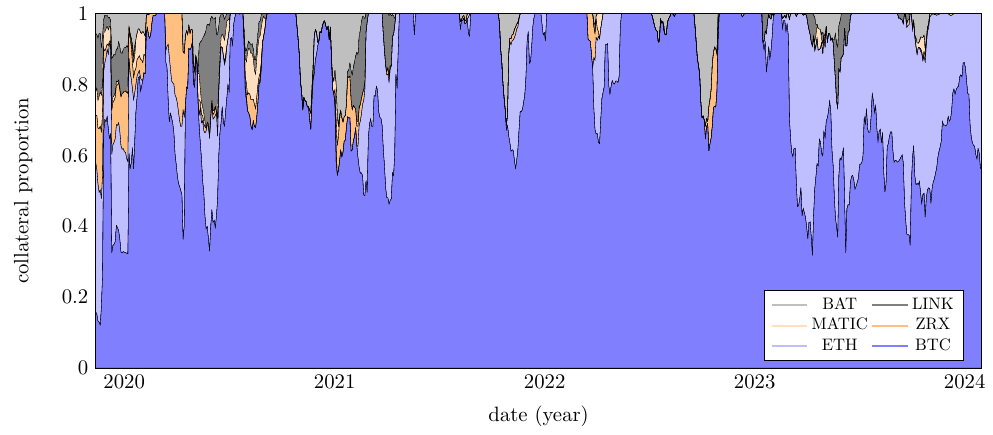}\label{fig:area_plot_optimal_30days_past}} \\
   \subfloat[][Optimal Portfolio: 60-Day Window]{ \includegraphics[width=1.0\linewidth]{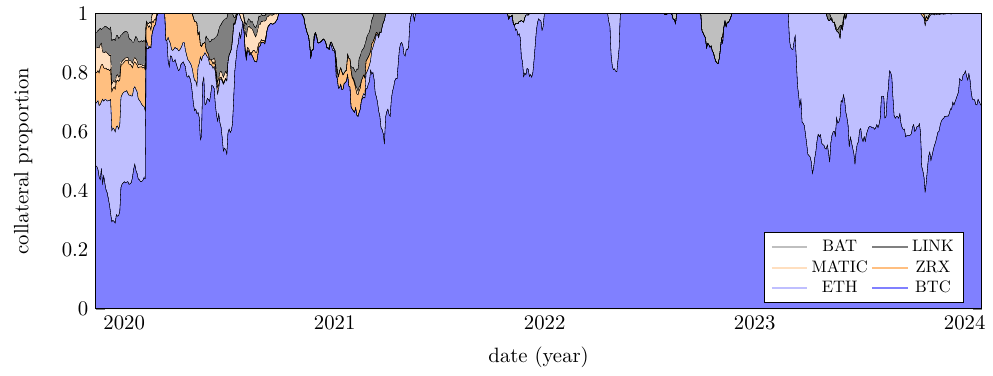}\label{fig:area_plot_optimal_60days_past}} \\
   \subfloat[][Optimal Portfolio: 200-Day Window]{ \includegraphics[width=1.0\linewidth]{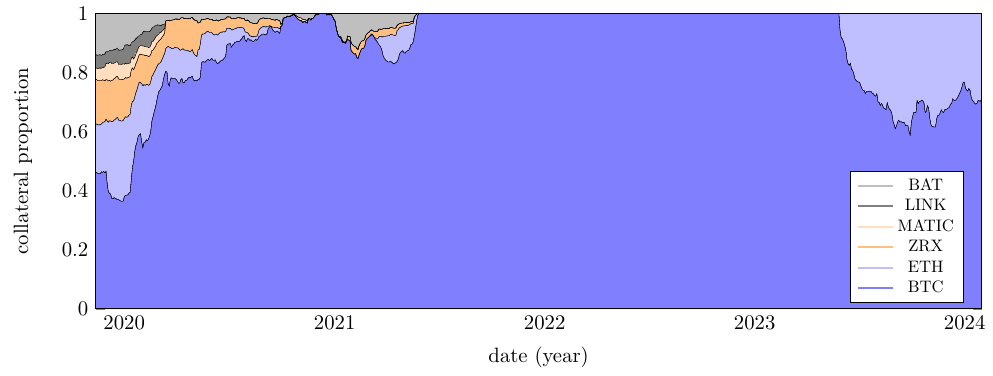}\label{fig:area_plot_optimal_200days_past}}
    \caption{Evolution of the optimal portfolios based on a 30-day window (top), 60-day (middle), or 200-day (bottom) of information.}
    \label{fig:area_plot_optimal}
\end{figure}

\textbf{Optimal Portfolios. }
The evolution of the optimal (i.e., semivariance-minimizing) portfolio over time is depicted in \Cref{fig:area_plot_optimal}, for a 30-day (\ref{fig:area_plot_optimal_30days_past}), 60-day (\ref{fig:area_plot_optimal_60days_past}), and 200-day (\ref{fig:area_plot_optimal_200days_past}) past information window to calculate the model parameters. The main observation is that the optimal portfolio is dominated by (W)BTC, although ETH is also included after mid-2023. This is largely due to BTC being less volatile than ETH, and the inclusion of ETH not having much added benefit due to the high correlation between their prices. We observe sharp changes in the optimal portfolio in \Cref{fig:area_plot_optimal_30days_past}. This is due to periods in which the changes in returns and/or volatility were uneven between BTC and ETH, which can have a major impact on portfolio composition for short inclusion windows. We observe that ERC-20 tokens are included in the 60-day and 200-day optimal portfolio before mid-2021, but not after. In general, the 200-day window portfolio seems to be a better choice than the 30-day and 60-day portfolios, which exhibit strong day-to-day changes in their composition. Research into appropriate window sizes is an interesting direction for future work.

\begin{table}[t]
    \centering
    \addtolength{\tabcolsep}{7pt}
    \begin{tabular}{lrrrrr}
    \toprule
     & DAI & A-Vol & A-Sem  & DAI-Vol & DAI-Sem \\
    \midrule
    Annual Volatility      & 79.82\% & 70.28\% & 70.30\% & 70.70\% & 70.74\% \\
    Annual Semideviation   & 57.00\% & 50.81\% & 50.69\% & 52.00\% & 51.96\% \\
    Historical Simulations      & 53.29\% & 48.47\% & 48.64\% & 55.08\% & 54.65\% \\
    GBM Simulations        & 61.90\% & 52.60\% & 54.60\% & 57.50\% & 59.20\% \\
    \bottomrule
    \end{tabular}
    \addtolength{\tabcolsep}{-7pt}
    \caption{Evaluation of average Dai portfolio and optimized portfolios.}
    \label{tab:portfolio_comparison}
\end{table}

\textbf{Comparison. }
\Cref{tab:portfolio_comparison} displays the various risk measures discussed in \Cref{sec:portfolio_optimization} for different portfolios. The Dai portfolio represents an average (across all historical data) historical collateral composition reduced to the six tokens with the highest contribution (BAT, LINK, MATIC, ZRX, ETH, and WBTC). DAI-Vol and DAI-Sem represent portfolios produced by optimal selection of the same six tokens using the optimization methods described in \Cref{sec:portfolio_optimization}. Finally, A-Vol and A-Sem are optimal portfolios selected from the full set of 38 tokens that were produced from the top 100 cryptocurrencies as described in \Cref{sec:portfolio_optimization}.

We observe from \Cref{tab:portfolio_comparison} that the A-Vol portfolio achieves a significant reduction of the annual portfolio volatility compared to the Dai average portfolio over time. The same applies to the annual semideviation, and portfolios produced by semivariance minimization  (A-Sem and DAI-Sem). All optimized portfolios perform better than the unoptimized Dai portfolio in most metrics. Compared to optimized Dai portfolios, we do find that the Dai portfolio yields better results when evaluated on historical price data. The reason is that optimization methods consider only the volatility, whereas some more volatile assets (such as ETH, which comprises a major part of Dai) had higher returns compared to the less volatile ones. This relates to the fact that for accurate estimation of expected returns, better portfolios may exist than those that minimize volatility. In general, the portfolios that were allowed to use non-Ethereum tokens performed better than the Ethereum-only portfolios,\footnote{This is an argument in favor of a greater diversity of wrapped tokens on the Ethereum blockchain, or integrated cross-chain support as in \cite{reijsbergen2023crocodai}.} but the difference was not as large as between the historical and optimized Dai portfolios (e.g., for the annual semideviation, the difference between DAI and DAI-Sem is roughly 5\%, but the difference between A-Sem and DAI-Sem is only 1.3\%). Although the failure probabilities are considerably high for all portfolios (i.e., between 48\% and 62\% in all cases), we emphasize that the simulations cover a relatively large period (365 days) and consider only the crypto part of the portfolio -- the main focus of our work is on the relative improvement. 
\rda{Protocol engineers can choose to either target the DAI-Sem or DAI-Vol portfolio, depending on the observed asymmetry between upward and downward price movements.}

\section{Conclusions \& Future Work}
\label{sec:conclusions}
In this work, we have presented a methodology for optimizing collateral portfolios through (semi)variance minimization. We have found that the optimal portfolios include substantially more (wrapped) Bitcoin than the actual portfolios, although the small contribution of tokens beyond ETH and BTC is shared between the actual and optimized portfolios.
\rda{Enforcing optimal portfolios is more naturally suited for hybrid fiat/crypto-backed stablecoins, in which the developer already controls a large portion of the collateral, than for pure crypto-backed stablecoins. For the former, transferring funds from real-world assets into crypto assets reduces the reliance on trusted third parties such as Monetalis, Coinbase, and BlockTower. (Although in the specific case of WBTC, there is another trusted third party in the form of the consortium that acts as bitcoin custodians.) For pure crypto-backed stablecoins, practical and legal challenges may hinder developers from using their own funds to buy collateral. For example, MakerDAO paid $\$1.5$ million in stamp duties to the Swiss government in 2022--2023 \cite{dlnews2023makerdao}. Such challenges are a limitation of the current work.}

\rda{As mentioned before in \Cref{sec:enforcing_portfolio}, alternative means of enforcing the optimal portfolio in pure crypto-backed stablecoins are an interesting direction for future work.} Another direction is the duration of the window that determines the recency threshold for data to be included in the computation of (co)variances -- there is a trade-off between including older data which may be irrelevant, and over-relying on a small set of recent data which may be subject to short-term jumps. 
A final interesting direction for future work is to take into account the costs -- particularly gas costs, but possibly also subsidies as mentioned above -- of making changes to the collateral portfolio. Our current method gives the optimal portfolio at each point in time given a recent history of prices, but it does not indicate when the difference between the current and optimal portfolios is large enough to warrant a change. An extension to our work would hence not only give a snapshot of the current optimal portfolio, but also indicate the right time to enact the change. To this end, future work could explore a meaningful metric to quantify the difference between the current and optimal portfolios. \rda{Such a metric could} take into account the various performance criteria (as displayed in \Cref{tab:portfolio_comparison}) and the costs of enacting the change, \rda{and could be useful for protocol engineers when they determine the magnitude of the subsidies.}

\section*{Acknowledgements}
This work was supported by Ministry of Education (MOE) Singapore’s Tier 2 Grant Award No. MOE-T2EP20120-0003. 

\bibliographystyle{plain}
\bibliography{ref}

\end{document}